\documentclass{emulateapj}

\usepackage{upgreek,gensymb,multirow}

\newcommand{\teff}{{$T_\mathrm{eff}$}}
\newcommand{\logg}{{log~{\em g}}}
\newcommand{\msun}{M$_{\odot}$}
\newcommand{\rsun}{R$_{\odot}$}
\newcommand{\lsun}{L$_{\odot}$}

\newcommand{\kms}{\mbox{km s$^{-1}$}}

\slugcomment{Submitted to Astrophysical Journal Letters}

\shorttitle{SN 2012aw}
\shortauthors{Fraser et al.}

\begin{document}

\title{Red and dead: The progenitor of SN 2012aw in M95}

\author{M. Fraser\altaffilmark{1},
	J.R. Maund\altaffilmark{1,8},
	S.J. Smartt\altaffilmark{1},
	M.-T. Botticella\altaffilmark{2},
	M. Dall'Ora\altaffilmark{2},
	C. Inserra\altaffilmark{1},
 	L. Tomasella\altaffilmark{3},\\
	S. Benetti\altaffilmark{3},
	S. Ciroi\altaffilmark{3},
	J.J. Eldridge\altaffilmark{4},
	M. Ergon\altaffilmark{5},
	R. Kotak\altaffilmark{1},
	S. Mattila\altaffilmark{6},
 	P. Ochner\altaffilmark{3},
	A. Pastorello\altaffilmark{3},\\
	E. Reilly\altaffilmark{1},
	J. Sollerman\altaffilmark{5},
	A. Stephens\altaffilmark{7},
	F. Taddia\altaffilmark{5},
	S. Valenti\altaffilmark{3}
	}

\altaffiltext{1}{School of Mathematics and Physics, Queens University Belfast, Belfast BT7 1NN, N. Ireland; m.fraser@qub.ac.uk}
\altaffiltext{2}{INAF Osservatorio Astronomico di Capodimonte, Salita Moiariello 16, I-80131, Napoli, Italia}
\altaffiltext{3}{INAF Osservatorio Astronomico di Padova, vicolo dellÕOsservatorio 5, 35122 Padova, Italy}
\altaffiltext{4}{Department of Physics, University of Auckland, Private Bag 92019, Auckland, New Zealand}
\altaffiltext{5}{The Oskar Klein Centre, Department of Astronomy, AlbaNova, Stockholm University, 10691 Stockholm, Sweden}
\altaffiltext{6}{Tuorla Observatory, Department of Physics and Astronomy, University of Turku, V\"ais\"al\"antie 20, FI-21500 Piikki\"o, Finland}
\altaffiltext{7}{Gemini Observatory, 670 North Aohoku Place, Hilo, HI 96720, USA}
\altaffiltext{8}{Royal Society Research Fellow}

\begin{abstract}

Core-collapse supernovae (SNe) are the spectacular finale to massive stellar evolution. In this Letter, we identify a progenitor for the nearby core-collapse SN 2012aw in both ground based near-infrared, and space based optical pre-explosion imaging. The SN itself appears to be a normal Type II Plateau event, reaching a bolometric luminosity of 10$^{42}$ erg s$^{-1}$ and photospheric velocities of $\sim$11,000 \kms\ from the position of the H$\beta$ P-Cygni minimum in the early SN spectra. We use an adaptive optics image to show that the SN is coincident to within 27 mas with a faint, red source in pre-explosion HST+WFPC2, VLT+ISAAC and NTT+SOFI images. The source has magnitudes $F555W$=26.70$\pm$0.06, $F814W$=23.39$\pm$0.02, $J$=21.1$\pm$0.2, $K$=19.1$\pm$0.4, which when compared to a grid of stellar models best matches a red supergiant. Interestingly, the spectral energy distribution of the progenitor also implies an extinction of $A_V>$1.2 mag, whereas the SN itself does not appear to be significantly extinguished. We interpret this as evidence for the destruction of dust in the SN explosion. The progenitor candidate has a luminosity between 5.0 and 5.6 log L/\lsun, corresponding to a ZAMS mass between 14 and 26 \msun\ (depending on $A_V$), which would make this one of the most massive progenitors found for a core-collapse SN to date.

\end{abstract}


\section{Introduction}

That massive red supergiants explode as Type IIP (Plateau) supernovae (SNe) at the end of their lives has been clearly established (for example, \citealp{Van03}, \citealp{Sma04}, \citealp{Mat08}, \citealp{Fra11}), and confirmed by the disappearance of their progenitors (\citealp{Mau09}). What remains of great interest, however, is understanding how SN progenitor properties (eg. mass, radius, metallicity) correlate with the explosion characteristics (kinetic energy, synthesised $^{56}$Ni, luminosity etc). The continued identification of progenitors of nearby SNe, coupled with follow-up observations of the SNe, can help shed light on this link. In this Letter we present an analysis of the progenitor of SN 2012aw, which is the coolest, and probably the most massive progenitor of a Type IIP SN found thus far.

SN 2012aw in M95 was discovered independently by several amateur astronomers (\citealp{Fag12}), with a first detection on 2012 Mar 16.9. A non-detection at a limiting magnitude of $R$ $\gtrsim$ 20.7 (\citealp{Poz12}) on Mar 15.3 sets a rigorous constraint on the age of the SN at discovery of $<$1.6 days, and we adopt an explosion epoch of Mar 16.0 UT ($\pm$ 0.8 d). \cite{Mun12} obtained a spectrum of the SN on March 17.8, which showed a featureless blue continuum; subsequent spectra (\citealp{Ito12,Siv12}) suggested the SN was a young Type IIP, although this classification is yet to be confirmed by the emergence of a plateau in the lightcurve at 3-4 weeks after explosion. A candidate progenitor for SN 2012aw was identified in archival Hubble Space Telescope data by \cite{Eli12} and \cite{Fra12}, who both suggested that a faint red supergiant was the likely precursor star. 
 
The host galaxy of SN 2012aw, M95, is a barred and ringed spiral galaxy with an inclination of 55$\degree$. \cite{Fre01} measured a Cepheid-based distance to M95 of 10 Mpc (corresponding to $\upmu=30.0\pm0.09$ mag), which we adopt in all of the the following. This distance is consistent with that obtained from the Tip of the Red Giant Branch (\citealp{Riz07}). We also adopt a foreground (Milky Way) extinction of $E(B-V) = 0.028$ mag from \cite{Sch98}. Using the line strengths for the closest H{\sc ii} region reported in \cite{Mcc85} we calculate a metallicity of 12+log[O/H]  = 8.8$\pm$0.1 using the [O{\sc iii}]/[N{\sc ii}] relation of \cite{Pet04}. From the recent study of the radial metallicity gradient in M95 by \cite{Pil06}, at the position of SN 2012aw we estimate a metallicity of 12+ log [O/H]  = 8.6$\pm$0.2. Hence we suggest that the metallicity of the progenitor of SN 2012aw is approximately solar (for comparison, Milky Way H{\sc ii} regions have a typical metallicity of 12+ log [O/H]  = 8.7$\pm$0.3; \citealp{Hun09}), albeit with the caveat that metallicity calibrations have significant uncertainties (as discussed in this context by \citealp{Sma09}).

SN 2012aw was detected in X-rays with the Swift X-ray Telescope by \cite{Imm12}, at a luminosity of $L_X = 9.2\pm2.5 \times 10^{38}$ erg s$^{-1}$. This is at the upper end of the range of Type IIP X-ray luminosities ($0.16 - 4 \times10^{38}$ erg s$^{-1}$; \citealp{Che06}), although still much lower than the typical values seen for Type IIn SNe ($0.2 - 1.6 \times10^{41}$ erg s$^{-1}$; \citealp{Fox00}) which are interacting with significant amounts of circumstellar material. SN 2012aw was also detected in radio observations (\citealp{Yad12}; \citealp{Sto12}) with a flux at 22 GHz of 2$\times10^{25}$ erg s$^{-1}$ Hz$^{-1}$ at +1 week after explosion, rising to 4$\times10^{25}$ erg s$^{-1}$ Hz$^{-1}$ at +2 weeks. These values are similar to those of the sample of Type IIP SNe presented by \cite{Che06}.

\section{Supernova characterization and followup}

Our collaboration\footnote{http://graspa.oapd.inaf.it/index.php?option=com\textunderscore content\&vi\\ew=article\&id=68\&Itemid=93} commenced an intensive spectroscopic and photometric follow-up campaign for SN 2012aw immediately after the discovery. Full coverage of the SN will be published in a future paper, here we present a limited set of optical observations from first two weeks after explosion to characterize the initial SN evolution.

\begin{figure}
\epsscale{1.10}
\plotone{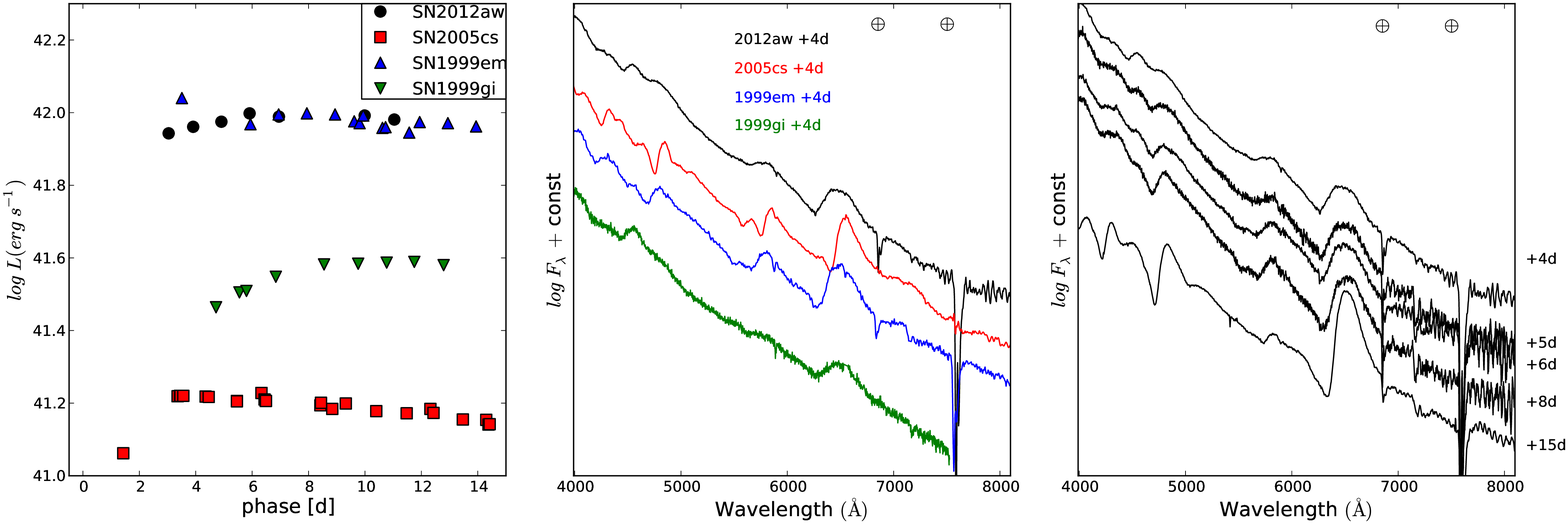}
\caption{Follow-up observations of SN 2012aw. Left panel: pseudo-bolometric ($BVRI$) lightcurve of SN 2012aw, as compared to other Type IIP SNe. Centre panel: comparison of early spectra of the same SNe to SN 2012aw (all at rest wavelength). Right panel: Sequence of spectra for SN 2012aw over the first two weeks since discovery.
\label{fig:followup}
}
\end{figure}

Photometry was obtained for SN 2012aw with the Asiago Observatory Schmidt telescope in the Landolt $BVRI$ system. The data were pipeline reduced, and PSF-fitting photometry was performed on the images using the QUBA pipeline (\citealp{Val11}). A pseudo-bolometric (ie. $BVRI$) lightcurve was constructed by integrating the observed SN flux over the optical filters. The resulting lightcurve is shown in Fig. \ref{fig:followup}, together with those of three other Type IIP SNe for comparison (SN 1999em, \citealp{Elm03}; SN 1999gi, \citealp{Leo02}; SN 2005cs, \citealp{Pas09}). The luminosity of SN 2012aw is comparable to that of the SN 1999em, placing it firmly in the continuum of normal Type IIP SNe, as opposed to sub-luminous events such as SN 2005cs.

Spectra of SN 2012aw were obtained from the Asiago 122cm telescope + Boller\&Chivens Spectrograph + 300tr/mm, and the Nordic Optical Telescope + ALFOSC + Gr\#4. Spectra were reduced, extracted, and flux and wavelength calibrated with the QUBA pipeline. The sequence of spectra from the first two weeks after explosion is presented in Fig. \ref{fig:followup}. Without any adjustment for either Milky Way or host galaxy extinction, the slope of the SN 2012aw spectrum (epoch 3.9 d) is almost identical to that of SN 1999em at +4 d \citep{Bar00,Ham01}. \citeauthor{Bar00} found that a synthetic spectrum fit with $T_{\rm BB}=11000-13000$\,K and an extinction of $E(B-V)\sim0.05$ mag provides a good match to the continuum slope and H{\sc i} lines. A similar result would necessarily be determined for SN 2012aw given its striking similarity to SN 1999em. The H$\alpha$ absorption minimum is at a velocity of $v_{phot}\sim13000$ \kms, again almost identical to the minimum for SNe 1999em and 1999gi. Fig. \ref{fig:followup} shows the early spectra of these three Type IIP SNe illustrating the similarity, also shown is the low energy, faint Type IIP SN 2005cs (\citealp{Pas09}) which has a H$\alpha$ minimum at a much lower velocity of  6500 \kms. This clearly shows that SN 2012aw is not a low energy explosion. 

A high resolution spectrum of SN 2012aw was obtained with the Telescopio Nazionale Galileo + Sarg on Mar 29. The two components of the Na{\sc i} D doublet were detected at 5907.998, 5914.004 \AA\ , which is consistent with the recessional velocity of M95. The equivalent width (EW) of the two components were 0.286 and 0.240 \AA\ respectively. From the empirical relation of \cite{Mun97}, we estimate E(B-V) = 0.10$\pm$0.05 mag for SN 2012aw. We note however, that even if dust has been photo-evaporated in the explosion (see Sect. \ref{s_results}), the Na{\sc i} absorption should remain unchanged and hence these calibrations may systematically over-estimate the extinction. 

\section{Progenitor observations and data analysis}

\subsection{Archival data}

\begin{figure}
\includegraphics[angle=0, width=43mm]{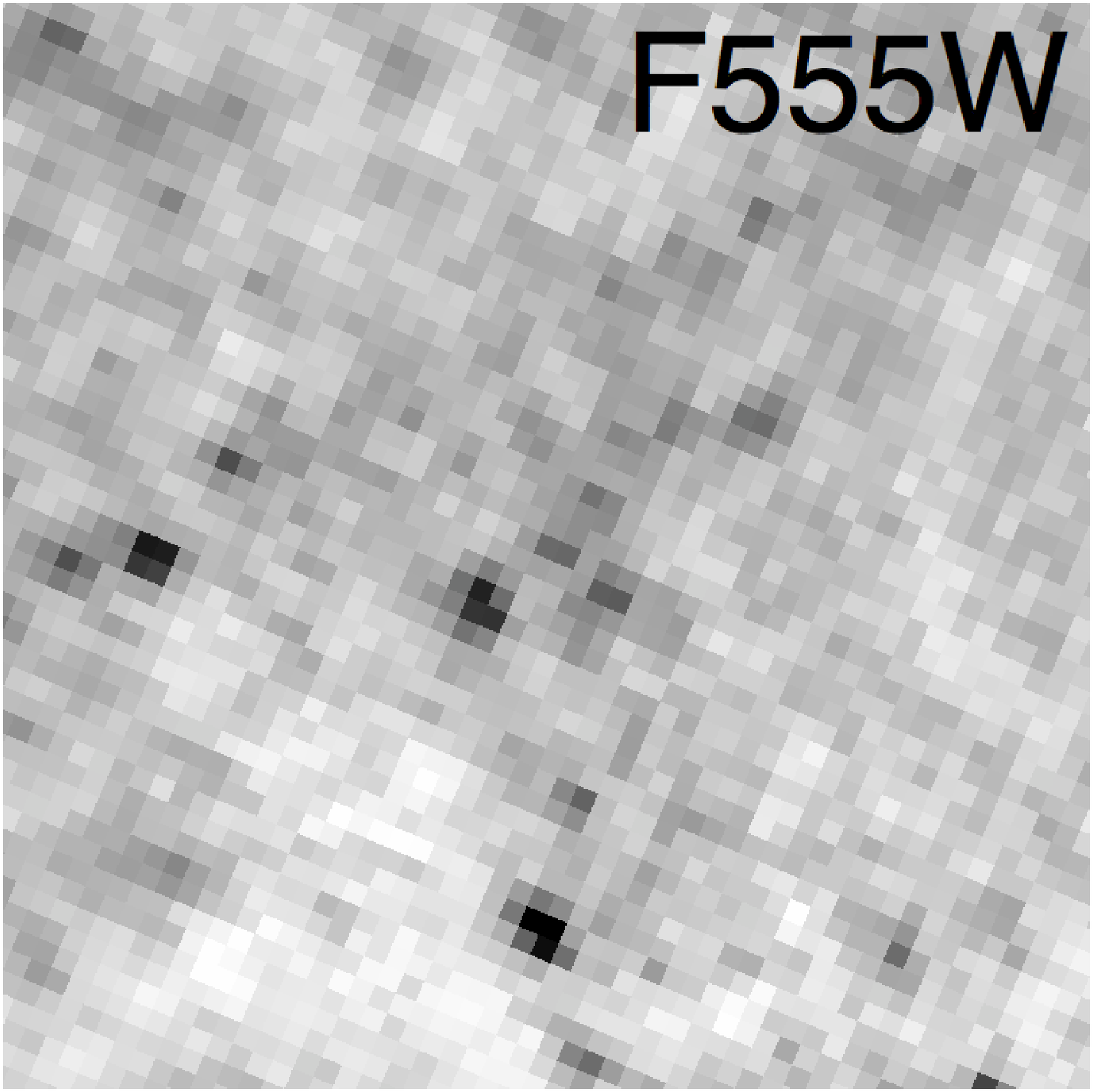}
\includegraphics[angle=0, width=43mm]{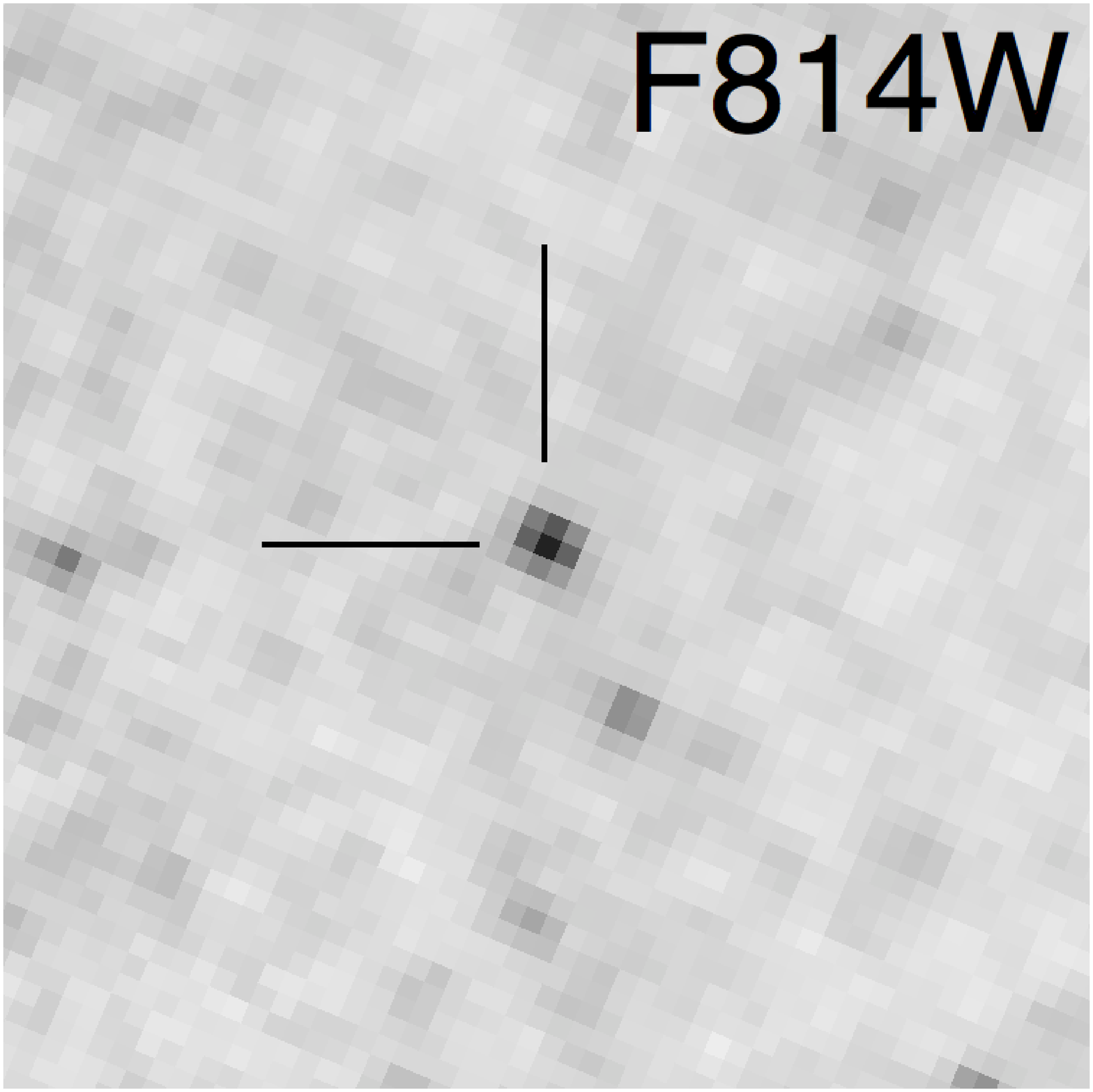}
\includegraphics[angle=0, width=43mm]{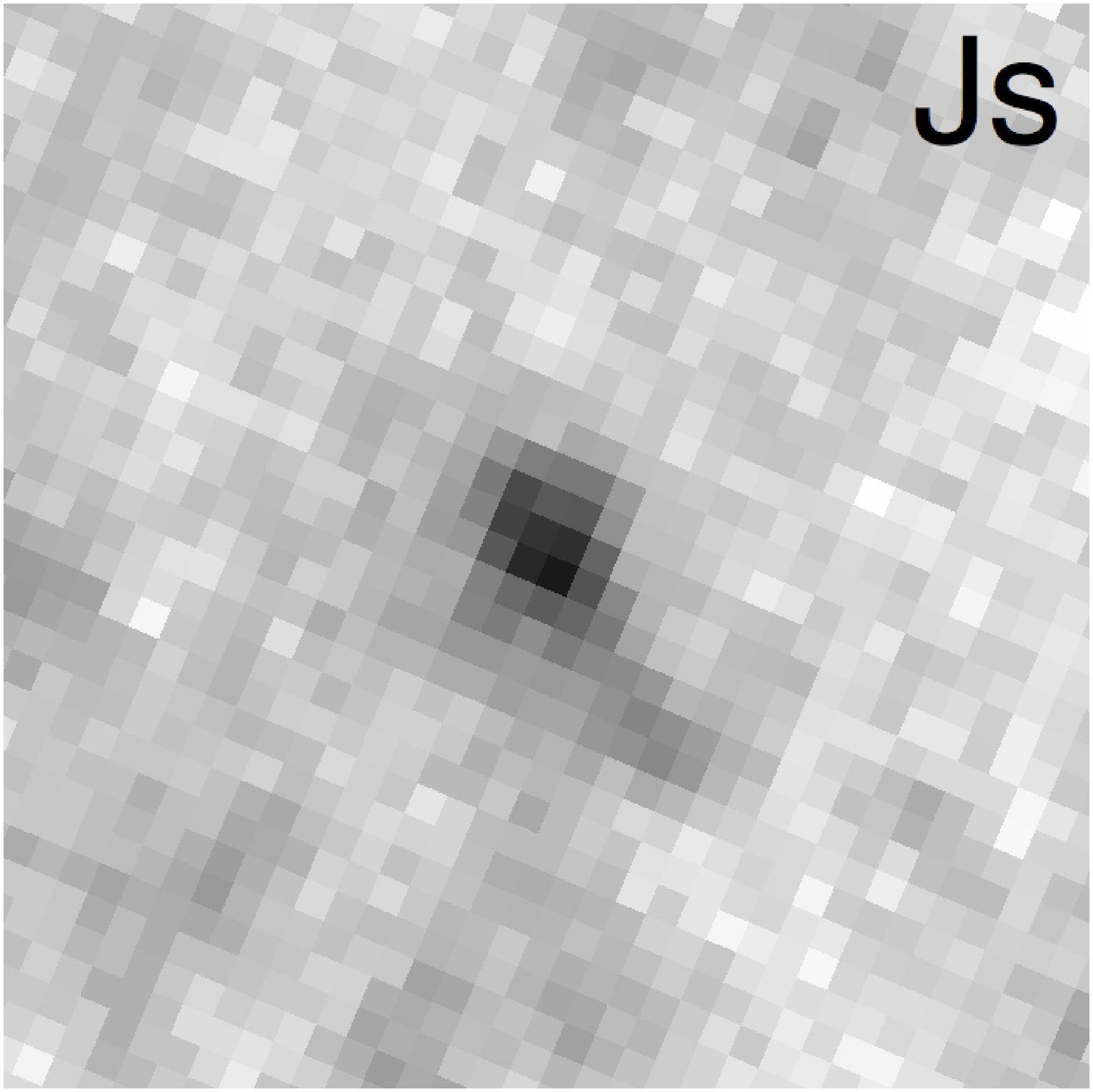}
\includegraphics[angle=0, width=43mm]{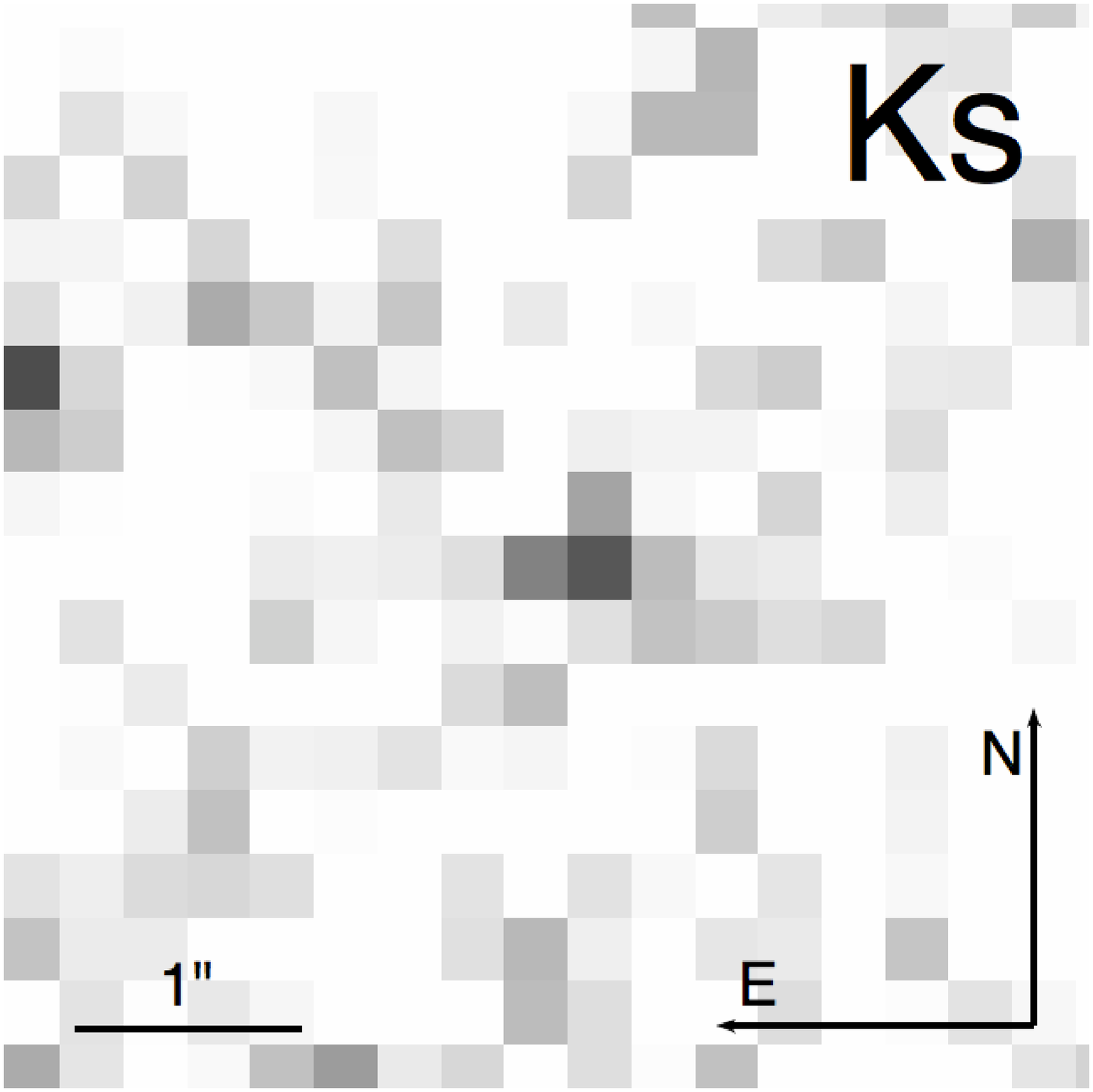}
\caption{The progenitor of SN 2012aw as detected in archival pre-explosion images. Images are scaled and oriented as indicated, each panel is centred on the progenitor position as indicated with tick marks in the $F814W$ image. In the $J$-band image the progenitor is partially blended with a nearby source, this has been corrected for with PSF-fitting photometry.
\label{fig:progenitor}
}
\end{figure}

We have searched both the Hubble and Spitzer Space Telescope archives, along with the publicly available archives of all the major ground based observing facilities, for pre-explosion images covering the site of SN 2012aw. A log of all data used is given in Table \ref{tab:pre}. The deepest, and highest resolution images of M95 found were from the Wide-Field and Planetary Camera 2 (WFPC2) onboard the Hubble Space Telescope (HST). The site of SN 2012aw was observed on multiple occasions between 1994 Nov and 1995 Jan with the $F439W$, $F555W$ and $F814W$ filters, and again in the $F555W$ filter in 1995 Dec, as part of the HST Key Project on the Extragalactic Distance Scale. HST +WFPC2 observed M95 again in $F336W$ and $F658N$ on 2009 Jan 18. M95 was also observed with the Subaru Telescope+ SuprimeCAM in $R$ on 1999 Jan 28, but as the seeing was $\gtrsim$ 2.5" in these images, they were of no use for constraining the progenitor. NTT+SOFI ($K_S$) and VLT+ISAAC ($J_S$) images of M95 were obtained from the ESO archive. A subset of the data used is shown in Fig. \ref{fig:progenitor}.

All HST data were downloaded from the Multimission Archive at STScI\footnote{http://archive.stsci.edu}. The individual WFPC2 $F814W$ images were combined using the drizzle algorithm within {\sc iraf}\footnote{IRAF is distributed by the National Optical Astronomy Observatories, which are operated by the Association of Universities for Research in Astronomy, Inc., under cooperative agreement with the National Science Foundation.} (\citealp{Fru02}), although as the offsets between the individual exposures were not in an optimal subpixel dither pattern, we did not resample the image to a smaller pixel scale. Small image shifts were removed using the {\sc tweakshifts} task to improve the quality of the alignment.

The individual VLT+ISAAC images were reduced (bad-pixel masked, flat-fielded and sky-subtracted) using the ISAAC pipeline (version 6.0.6). The individual frames from each night where the SN position was in the field of view were then aligned with sub-pixel shifts and combined within {\sc iraf}. The image from the second night of ISAAC observations (2000 Mar 27) was used for all of the subsequent analysis, as the total on source exposures are deeper, and the SN position is further from the edge of the chip. The NTT+SOFI data from 2006 Mar 24 were not flat-fielded (as appropriate calibrations were not available), but were sky subtracted with the SOFI pipeline (version 1.5.4) before the individual on-source frames were combined. A total of 600s was obtained on source, with a FWHM of 0.6\arcsec. For the SOFI data from 2002 Mar 25, the SOFI pipeline was unable to reduce the data, and so the {\sc iraf xdimsum} package was used for the reduction. 

The Spitzer Space Telescope + IRAC observed M95 as part of the SINGS survey (\citealp{Ken03}), however the spatial resolution of these data (1.7\arcsec\ PSF FWHM in the shortest wavelength channel) is ill suited for progenitor identification. We downloaded the IRAC data from the Spitzer Heritage Archive, but could not identify a point source at the SN coordinates. We have not considered the IRAC data any further in the following.

\subsection{Progenitor identification}

A deep, high resolution $K$-band image of SN 2012aw was obtained on 2012 Mar 31 with Gemini+NIRI (Program: GN-2012A-Q-38). To match the resolution of the pre-explosion WFPC2 images, we used the ALTAIR adaptive optics system to correct for the seeing; as the SN was sufficiently bright (mag$\sim$13) at the time of the observations it was used as a natural guide star for ALTAIR. The f14 camera was used, which has a 0.05\arcsec\ pixel scale over a 51\arcsec$\times$51\arcsec\ field of view. As the region around SN 2012aw is not crowded, we used on source dithers of a few arcseconds, and median combined these to create the sky frame for each exposure. The data were reduced with the {\sc iraf gemini} package, yielding a final combined image with a PSF FWHM of 0.2\arcsec.

40 sources were identified common to both the NIRI and the drizzled WFPC2 $F814W$ images. The positions of these reference stars were measured in both images with {\sc iraf phot}. After rejecting outliers from the fit, which either had low signal to noise or centering errors, we used 22 sources to derive a geometric transformation between the pixel coordinate systems of the WFPC2 and NIRI images. A general fit allowing for scaling, a shift in x and y, rotation and a skew term was used, with a root mean square error in the fit of 53 mas. The pixel coordinates of SN 2012aw were measured in the NIRI image, and transformed to the pixel coordinates of the WFPC2 image. The transformed position of the SN coincided with the same source identified by \citealp{Eli12} and \citealp{Fra12}. The position of this source was measured using the three different centering algorithms in {\sc phot}, all of which agreed to within 14 mas. The separation between the transformed SN position, and the progenitor candidate was 27 mas, which is well within the total (transformation + SN position + progenitor position) uncertainty of 55 mas. Hence we formally identify the source as the progenitor candidate for SN 2012aw. 

\subsection{Progenitor photometry}

Photometry was performed on the WFPC2 images with the {\sc hstphot} package (\citealp{Dol00}), which is designed specifically for undersampled data from this instrument. The pipeline reduced images were first masked for bad pixels, hot pixels and cosmic rays using the ancillary packages distributed with {\sc hstphot}. After masking, PSF-fitting photometry was performed on the individual images, which typically had exposure times of 1000 to 2000 s. The SN progenitor candidate was detected at a magnitude of $F555W$=26.70$\pm$0.06 and $F814W$=23.39$\pm$0.02 (in the photometric system as defined in \citealp{Dol00b}), but was not detected in either the $F336W$ or $F439W$ filter images. Based on photometry of sources detected by {\sc hstphot} in the $F439W$ image, we estimate a 5$\sigma$ limiting magnitude of $F439W>$26. The $F814W$ magnitude for the progenitor, as measured in the individual frames, appears to be unchanged within the uncertainties over the $\sim$50 day period from the first to the last observation. Unfortunately the progenitor is too faint to check for variability in the $F555W$ images. The $F658N$ (H$\alpha$) WFPC2 image was also examined, but there was no sign of emission at the SN position.

The SOFI $K_S$ and ISAAC $J_S$ images were aligned to an $F814W$-filter mosaic of the four WFPC2 chips, and a counterpart to the optical progenitor candidate was found in both images. PSF-fitting photometry was performed on the ISAAC $J_\mathrm{s}$ and SOFI $K_\mathrm{s}$ image with the {\sc snoopy} package in {\sc iraf}. A magnitude was found for the progenitor of $J$ = 21.1$\pm$0.2 (setting the photometric zeropoint from the $J$ magnitude of two 2MASS sources in the field, which is the dominant source of error). As the bandpasses of the $J_\mathrm{s}$ and $J$ filters are different, we calculated colour terms from synthetic photometry of MARCS model spectra between 3000 and 4000 K. Over this temperature range the colour term is $\lesssim$ 0.02 mag, which is significantly less than our photometric error. For the SOFI K$_\mathrm{s}$ image from 2006, we measured a magnitude of $K$ = 19.3 $\pm$ 0.4 with PSF-fitting photometry. The uncertainty in the measurement consists of 0.36 mag from the progenitor photometry, and 0.21 uncertainty in the zero point, which was set from seven 2MASS sources in the field. As a check of the PSF-fitting measurement, aperture photometry was performed on the same image with {\sc iraf} phot, which returned a magnitude of $K$ = 19.1$\pm$0.4. Aperture photometry of the SOFI image from 2002 yielded a magnitude for the progenitor of 18.9$\pm$0.3 mag. We hence adopt an average value of $K$=19.1 from the two epochs for the progenitor magnitude, with a conservative error of $\pm$0.4 mag. All NIR magnitudes are in the 2MASS system (\citealp{Coh03}).

\section{Results and Discussion}
\label{s_results}

\begin{figure}
\epsscale{1.10}
\plotone{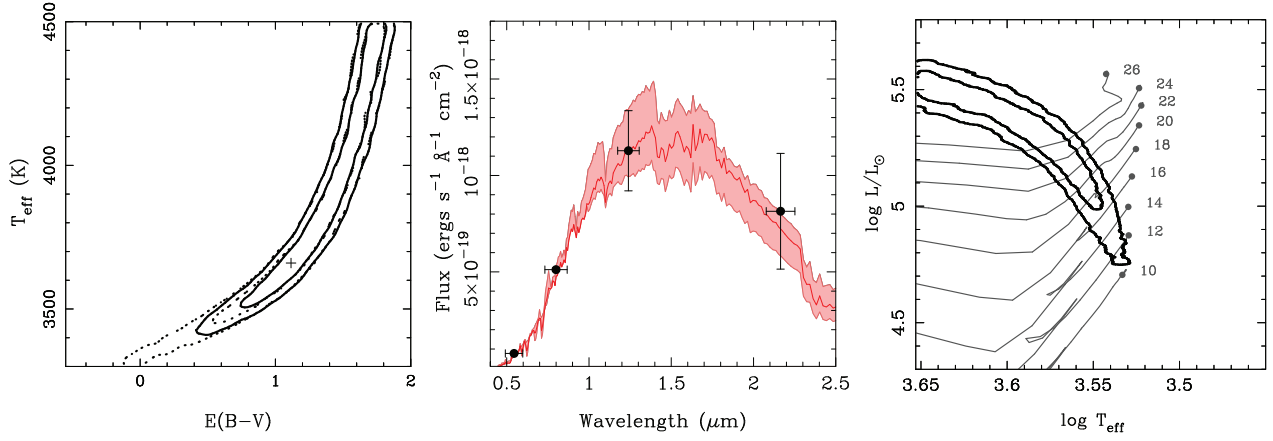}
\caption{Left panel: posterior probability distribution for reddening and temperature for solar metallicity MARCS SEDs (Contours are shown for 68\% and 95\% credibility intervals).  The inclusion of the $K$-band observation (solid contours) provides a significant constraint on the possible values of reddening, over the fits that only utilise the observed $VIJ$ fluxes (dotted contours).  Centre panel: Best fit SED (dark red line) for \teff=3660 K and E(B-V)=1.12), as compared to the observed progenitor magnitudes. The shaded region indicates the allowed range of spectra (corresponding to the 68\% region in the left panel) with an upper bound of \teff=4490 K, E(B-V)=1.758 mag, and a lower bound of \teff=3500, E(B-V)=0.74 mag. Right panel: Hertzprung-Russell diagram showing the final probability contours for the progenitor of SN 2012aw, together with the STARS evolutionary tracks (labelled with ZAMS mass in \msun) as discussed in the text.
\label{fig:sed}
}
\end{figure}

We used our own Bayesian SED-fitting code (based on the Bayesian Inference-X nested sampling framework; Maund 2012 in prep; \citealp{Ski04}) to compare the observed progenitor magnitudes to synthetic photometry of MARCS model spectra (\citealp{Gus08}). The models used were for 15 \msun\ stars with \logg\ = 0 dex and solar metallicity, with temperatures between 3300 and 4400 K. The SEDs were rebinned by a factor of 10 prior to the calculation of synthetic photometry (\citealp{Ple08}). We fit the progenitor $VIJK$ magnitudes with flat priors on \teff\ and E(B-V), which was allowed to vary between $-0.5 <$ E(B-V) $< 2$. A match was found with the observed progenitor magnitudes along a banana-shaped region as shown in Fig. \ref{fig:sed}.

The luminosity is ill constrained due to the uncertain extinction, and indeed varies within the 1$\sigma$ contours between 5.0 dex at 3550 K to 5.6 dex for a progenitor with \teff=4450 K. The SED fit is poorer in $K$ than the $J$ or WFPC2 data, as can be seen in Fig. \ref{fig:sed}. Besides the larger photometric errors in $K$ data, the models are also quite sensitive to metalliciity at cooler temperatures, due to the TiO absorption which is present in the NIR spectra of cool stars. Determining the luminosity of the progenitor from the $K$-band magnitude only (as A$_K\sim$0.1 A$_V$, and the bolometric correction to $K$-band only changes by 1.1 mag between 3300 and 4500 K) gives a value between 4.9 and 5.5 dex. We also attempted to fit the progenitor SED with MARCS spectra of twice solar metallicity models, but found the fit to be poorer than for the solar metallicity models.

In Fig. \ref{fig:sed}, we have also plotted stellar evolutionary tracks from the STARS code (\citealp{Eld08}). Comparing the luminosity of the progenitor to the endpoints of the tracks implies a ZAMS mass of between 14 and 26 \msun. If we consider the hotter progenitors in Fig. \ref{fig:sed}, the discrepancy in temperature with the end points of the STARS evolutionary tracks becomes more apparent. However, increased mass loss, either in a binary or through rotation, would serve to bring the end points of these tracks over to hotter temperatures (\citealp{Geo12}).

As the radius of the progenitor can be expressed as R=$\sqrt{L/T_{eff}^4}$, it is easy to calculate the expected radius for each point in the HR diagram. We find that even for a temperature of 4500 K, and a luminosity of 5 dex, the radius of the progenitor is still $>$500 \rsun. This is sufficiently large that we may reasonably expect the progenitor to give rise to a Type IIP SN (e.g.. \citealp{Pop93,Kas09}), and so cannot be used to further restrict the region of the HRD where the progenitor lies.

Within the 68\% confidence contours, the solar metallicity SEDs favour an extinction which is greater than $E(B-V)>0.8$ mag, although within 95\% (2$\sigma$), there is a solution with a correspondingly poorer fit for $E(B-V)=0.4$ mag. The implication of this is that a significant amount of dust could have been destroyed in the initial phases of the SN explosion (eg. \citealp{Wax00,Dwe08}), and that the extinction towards a SN can \emph{not} be taken as a proxy for the extinction towards the progenitor. We stress, however, that regardless of extinction, SN 2012aw potentially arises from one of the highest mass progenitors found to date.

As was recently suggested by \cite{Wal11}, significant amounts of circumstellar dust around SN progenitors is an appealing solution to the lack of high mass red supergiant progenitors identified by \cite{Sma09}. We caution however, that SN 2012aw is the \emph{reddest} SN progenitor found thus far, and appears to suffer from extinction that is comparable to that of the most luminous Galactic red supergiants (\citealp{Lev05}), but higher than is typical for a Type IIP progenitor (from the limited sample with colour information). Hence it remains unclear whether circumstellar dust truly is the panacea for the ``red supergiant problem''.

\acknowledgments

We acknowledge funding from STFC (MF), the ERC (SJS) and the Royal Society (JRM). SB, MT, IJD and AP are partially supported by the PRIN-INAF 2009 with the project ``Supernovae Variety and Nucleosynthesis Yields''.

Based on observations obtained at the Gemini Observatory, which is operated by the Association of Universities for Research in Astronomy, Inc., under a cooperative agreement with the NSF on behalf of the Gemini partnership.

Based on observations made with the NASA/ESA Hubble Space Telescope, obtained from the data archive at the Space Telescope Science Institute. STScI is operated by the Association of Universities for Research in Astronomy, Inc. under NASA contract NAS 5-26555. Based on data obtained from the ESO Science Archive Facility. Partially based on observations collected at Asiago observatory, Galileo 1.22 and Schmidt 67/92 telescopes operated by Padova University and INAF OAPd, and on observations made with the NOT, operated on the island of La Palma jointly by Denmark, Finland, Iceland, Norway and Sweden, in the Spanish Observatorio del Roque de los Muchachos of the Instituto de Astrofisica de Canarias.
 
We thank Ben Davies and Rolf Kudritzki for useful advice on metallicities.
 
Facilities: Gemini(NIRI+ALTAIR), HST(WFPC2), VLT(ISAAC), NTT(SOFI)

\clearpage

\clearpage

\clearpage

\clearpage

\begin{deluxetable}{llllc}
\tabletypesize{\scriptsize}
\tablecaption{Pre-explosion images of the site of SN 2012aw
\label{tab:pre}}
\tablewidth{0pt}
\tablehead{
\colhead{Date} &
\colhead{Telescope} & 
\colhead{Instrument} &
\colhead{Filter} &
\colhead{Exposure (s)}
}
\startdata
2009-01-18	& HST	& WFPC2			& F336W	&	4400		\\
1994-11-29	& \multirow{2}{*}{-}	& \multirow{2}{*}{-}	&  \multirow{2}{*}{F439W}		&  \multirow{2}{*}{5000}			\\
--1994-12-19  	&		&				&		&			\\
2009-01-18	& -		& -				& F658N	&	1800		\\
1994-11-29	& \multirow{2}{*}{-}	& \multirow{2}{*}{-}	&  \multirow{2}{*}{F555W}		&  \multirow{2}{*}{34130}			\\
--1995-12-04  	&		&				&		&			\\
1994-11-29	& \multirow{2}{*}{-}	& \multirow{2}{*}{-}	&  \multirow{2}{*}{F814W}		&  \multirow{2}{*}{9830}			\\
--1995-01-16  	&		&				&		&			\smallskip\\
2000-03-26	& VLT	& ISAAC			& J		& 	2400		\\   
2000-03-27	& -		& -				& -		&      4080		\smallskip\\   
2002-03-25	& NTT	& SOFI			& Ks 		&	600	 	\\	
2006-03-24	& -		& -				& - 		&	600	 	\\	
\enddata
\end{deluxetable}


\begin{thebibliography}{}

\bibitem[Baron et al.(2000)]{Bar00}
	Baron E. et al., 2000, ApJ, 545, 444

\bibitem[Chevalier et al.(2006)]{Che06}
	Chevalier R.A., Fransson C., Nymark T.K., 2006, ApJ, 641, 1029

\bibitem[Cohen et al.(2003)]{Coh03}
	Cohen M., Wheaton Wm.A., Megeath S.T., 2003, AJ, 126, 1090

\bibitem[Dolphin(2000a)]{Dol00}
	Dolphin A.E., 2000a, PASP, 112, 1383

\bibitem[Dolphin(2000b)]{Dol00b}
	Dolphin A.E., 2000b, PASP, 112, 1397

\bibitem[Dwek et al.(2008)]{Dwe08}
	Dwek E. et al., 2008, ApJ, 676, 1029

\bibitem[Eldridge et al.(2008)]{Eld08}
	Eldridge J.J., Izzard R.G., Tout C.A., 2008, MNRAS, 384, 1109

\bibitem[Elias-Rosa et al.(2012)]{Eli12}
	Elias-Rosa N., Van Dyk S.D., Cuillandre J.C.,  Cenko S.B., Filippenko A.V., 2012, ATEL, 3991

\bibitem[Elmhamdi et al.(2003)]{Elm03}
	Elmhamdi A. et al., 2003, MNRAS, 338, 939

\bibitem[Fagotti et al.(2012)]{Fag12}
	Fagotti P. et al., 2012, CBET, 3054, 1

\bibitem[Fox et al.(2000)]{Fox00}
	Fox D.W. et al., 2000, MNRAS, 319, 1154

\bibitem[Fraser et al.(2011)]{Fra11}
	Fraser M., et al. 2011, MNRAS, 417, 1417

\bibitem[Fraser et al. (2012)]{Fra12}
	Fraser M., Maund J.R., Smartt S.J., Kotak R., Sollerman J., Ergon M., 2012, ATEL, 3994
	
\bibitem[Freedman et al.(2001)]{Fre01}
	Freedman W.L. et al., 2001, ApJ, 553, 47
	
\bibitem[Fruchter \& Hook(2002)]{Fru02}
	Fruchter A.S., Hook R.N., 2002, PASP, 114, 144

\bibitem[Georgy(2012)]{Geo12}
	Georgy C., 2012, A\&A, 538

\bibitem[Gustafsson et al.(2008)]{Gus08}
	Gustafsson B., Edvardsson B., Eriksson K., J\o rgensen U.G., Nordlund \AA., Plez B., 2008, A\&A, 486, 951
	
\bibitem[Hamuy et al.(2001)]{Ham01}
	Hamuy M. et al., 2001, ApJ, 558, 615
	
\bibitem[Hunter et al.(2009)]{Hun09}
	Hunter I. et al., 2009, A\&A, 496, 841

\bibitem[Immler \& Brown(2012)]{Imm12}
	Immler \& Brown, 2012, ATEL, 3995

\bibitem[Itoh et al.(2012)]{Ito12}	
	Itoh R., Ui T., Yamanaka M., 2012, CBET, 3054, 2

\bibitem[Kasen \& Woosley(2009)]{Kas09}
	Kasen D., Woosley S.E., 2009, ApJ, 703, 2205

\bibitem[Kennicutt et al.(2003)]{Ken03}
	Kennicutt R.C. et al., 2003, PASP, 115, 928

\bibitem[Leonard et al.(2002)]{Leo02}
	Leonard D.C. et al., 2002, ApJ, 124, 2490
	
\bibitem[Levesque et al.(2005)]{Lev05}
	Levesque E.M., Massey P., Olsen K.A.G., Plez B., Josselin E., Meynet G., Maeder A., 2005, ApJ, 628, 973

\bibitem[McCall et al.(1985)]{Mcc85}
	McCall M.L., Rybski P.M., Shields G.A., 1985, ApJS, 57, 1

\bibitem[Mattila et al.(2008)]{Mat08}
	Mattila S., Smartt S.J., Eldridge J.J., Maund J.R., Crockett R.M., Danziger I.J., 2008, ApJ, 688, 91

\bibitem[Maund \& Smartt(2009)]{Mau09}
	Maund J.R., Smartt S.J., 2009, Science, 324, 486

\bibitem[Munari \& Zwitter(1997)]{Mun97}
	Munari U., Zwitter T., 1997, A\&A, 318, 269
	
\bibitem[Munari(2012)]{Mun12}
	Munari U., 2012, CBET, 3054, 3	

\bibitem[Pastorello et al.(2009)]{Pas09}
	Pastorello et al., 2009, MNRAS, 394, 2266

\bibitem[Pettini \& Pagel(2004)]{Pet04}
	Pettini M., Pagel B.E.J., 2004, MNRAS, 348, 59

\bibitem[Pilyugin, Thuan \& V\'ilchez(2006)]{Pil06}
	Pilyugin L.S., Thuan T.X., V\'ilchez J.M., 2006, MNRAS, 367, 1139

\bibitem[Plez(2008)]{Ple08}
	Plez B., 2008, in Astronomical Spectroscopy and Virtual Observatory, Eds.: M. Guainazzi and P. Osuna, pp.169 

\bibitem[Popov(1993)]{Pop93}
	Popov D.V., 1993, ApJ, 414, 712

\bibitem[Poznanski et al.(2012)]{Poz12}
	Poznanski D., Nugent P.E., Ofek E.O., Gal-Yam A., Kasliwal M.M., 2012, ATEL, 3996

\bibitem[Rizzi et al.(2007)]{Riz07}
	Rizzi L., Tully B., Makarov D., Makarova L., Dophin A.E., Sakai S., Shaya E.J., 2007, ApJ, 661, 815

\bibitem[Schlegel et al.(1998)]{Sch98}
	Schlegel D.J., Finkbeiner D.P., Davis M., 1998, ApJ, 500, 525
	
\bibitem[Siviero et al.(2012)]{Siv12}	
	Siviero A. et al., 2012, CBET, 3054, 4

\bibitem[Skilling(2004)]{Ski04}
	Skilling J., 2004, in American Institute of Physics Conference Series, Vol. 735, American Institute of Physics Conference Series, R. Fischer, R. Preuss, U. V. Toussaint, ed., pp. 395 - 405
	
\bibitem[Smartt et al.(2004)]{Sma04}
	Smartt S.J., Maund J.R., Hendry M.A., Tout C.A., Gilmore G.F., Mattila S., Benn C.R., 2004, Science, 303, 499
	
\bibitem[Smartt et al.(2009)]{Sma09}
	Smartt S.J., Eldridge J.J., Crockett R.M., Maund J.R., 2009, MNRAS, 395, 1409	
	
\bibitem[Stockdale et al.(2012)]{Sto12}
	Stockdale C.J. et al., 2012, ATEL, 4012

\bibitem[Walmswell \& Eldridge(2011)]{Wal11}
	Walmswell J.J., Eldridge J.J., 2011, MNRAS, 419, 2054
	
\bibitem[Waxman \& Draine(2000)]{Wax00}	
	Waxman E., Draine B.T., 2009, ApJ, 537, 796
	
\bibitem[Valenti et al.(2011)]{Val11}
	Valenti S. et al., 2011, MNRAS, 416, 3138
	
\bibitem[Van Dyk et al.(2003)]{Van03}
	Van Dyk S.D., Li W., Filippenko A.V., 2003, PASP, 115, 1289

\bibitem[Yadav et al.(2012)]{Yad12}
	Yadav N., Chakraborti S., Ray A., 2012, ATEL, 4010
	
\end{thebibliography}
\end{document}